\documentclass[]{aa}
\usepackage{psfig}
\usepackage{natbib}
\begin{document}
\thesaurus{01
           (12.07.1;
            11.17.4;
            11.17.1)}
\title{Adaptive optics imaging and integral field spectroscopy of
APM~08279+5255: Evidence for gravitational lensing\thanks{Based on data
collected at the Canada-France-Hawaii Telescope supported by INSU at Mauna
Kea, Hawaii (USA).}}
\author{C\'edric Ledoux\inst{1}
   \and Bertrand Th\'eodore\inst{2} 
   \and Patrick Petitjean\inst{3,4}
   \and Malcolm N. Bremer\inst{3}
   \and Geraint F. Lewis\inst{5,6,}\thanks{Fellow of the Pacific Institute for Mathematical Sciences 1998-1999}
   \and Rodrigo A. Ibata\inst{7}
   \and Michael J. Irwin\inst{8}
   \and Edward  J. Totten\inst{9}}
\institute{
   $^1$ Observatoire Astronomique de Strasbourg, 11 Rue de l'Universit\'e, F--67000 Strasbourg, France\\
   $^2$ Service d'A\'eronomie du CNRS, BP 3, F--91371 Verri\`ere le Buisson, France\\
   $^3$ Institut d'Astrophysique de Paris - CNRS, 98bis Boulevard Arago, F--75014 Paris, France\\
   $^4$ DAEC, Observatoire de Paris-Meudon, F--92195 Meudon Principal Cedex, France\\
   $^5$ Department of Physics and Astronomy, University of Victoria, Victoria BC, Canada\\
   $^6$ Astronomy Department, University of Washington, Seattle WA, USA\\
   $^7$ European Southern Observatory, Karl-Schwarzschild Stra$\ss$e 2, D--85748 Garching bei M\"unchen, Germany\\
   $^8$ Royal Greenwich Observatory, Madingley Road, Cambridge CB3 0EZ, UK\\
   $^9$ Department of Physics, Keele University, Keele, Staffordshire ST5 5BG, UK}
\date{Received 1 September 1998 / Accepted 18 September 1998}
\offprints{C. Ledoux (cedric@astro.u-strasbg.fr).}
\authorrunning{C. Ledoux et al.}
\titlerunning{Adaptive optics imaging and integral field spectroscopy of APM~08279+5255}
\maketitle
\begin{abstract}
We report observations of the $z$~=~3.87
broad absorption line quasar APM~08279+5255 (Irwin et al. 1998)
with the Adaptive Optics Bonnette (AOB) of the 
Canada-France-Hawaii Telescope.
The object is found to be a double source. The separation of the two images is
0$\farcs$35~$\pm$~0$\farcs$02 and the intensity ratio $I_{\rm north}$/$I_{\rm
south}$ = 1.21~$\pm$~0.25 in the $H$-band. No other image is detected down to
$H$(5$\sigma$)~=~21.3 within 10$\arcsec$ from the double image.
Strong support for the lensing hypothesis comes from the uniformity of the 
quasar spectrum as a function of spatial position in the image obtained 
with the integral field spectrograph OASIS at CFHT. From the
2D-spectroscopy,
narrow-band images are reconstructed
over the wavelength range 5600-6200~\AA\ to search for
emission-line objects in a field of 15$\arcsec\times$12$\arcsec$ around the
quasar. We find no  such object to  a limit of 6$\times$10$^{-17}$ erg
cm$^{-2}$ s$^{-1}$. We use the images centered on the deepest
absorption lines of the Ly$\alpha$ forest to dim the quasar and to increase
the sensitivity closer to the line of sight. One of the images, centered at
5766.4~\AA , exhibits a 3$\sigma$ excess 1.5$\arcsec$ from the quasar to the
north-east. 
%
\end{abstract}
\section{Introduction}
Recently, \citet{1998ApJ...505..529I} reported on the discovery of a highly
luminous broad absorption line quasar, APM 08279+ 5255, at a redshift $z_{\rm
em}$~=~3.87, positionally coincident within one arc second with the IRAS FSC
source F08279+5255. The object has an apparent $R$-band magnitude of 15.2 and
an overall SED indicative of one of the most luminous objects of the universe,
even in the probable situation that it is gravitationally lensed.
\citet{1998ApJ...505..529I} favored this latter explanation on the basis of a
point-spread function (PSF) fitting of two discrete sources to an image of the
quasar. They found a best fit with two components separated by $\sim$
0$\farcs$4 and an intensity ratio $\sim$ 1.1. Several metal line systems are
present in the quasar spectrum which could arise in a lensing object: a strong
\ion{Mg}{ii} system at $z_{\rm abs}$~=~1.18, another \ion{Mg}{ii} system at
$z_{\rm abs}$~=~1.81 and a damped Ly$\alpha$ candidate at $z_{\rm
abs}$~=~3.07.

Sub-mm observations with SCUBA \citep{1998ApJ...505L...1L} indicate
that the object contains an amount of dust comparable to that of the most
luminous sub-millimeter sources at high-redshift
\citep[e.g.][]{1992ApJ...398L..25D,1996Natur.382..428O,1998.astroph9804257}.
This holds true even if the object is lensed. The sub-mm spectral energy
distribution is consistent with emission by an optically thick black-body at a
temperature of 220~K, with a minimum dust mass of
4$\times$10$^{9}$~M$_{\odot}$.

In this Letter, we report adaptive optics imaging and integral field
spectroscopy of the field of APM~08279+5255.
%
\section{Observations}
\subsection{Adaptive optics image}
\begin{figure}[t]
\flushleft{\hbox{
\psfig{figure=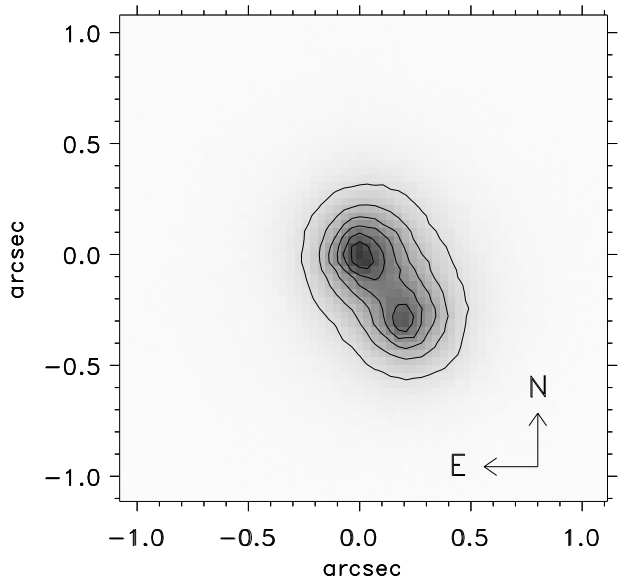,width=8.1cm,clip=,bbllx=18.pt,bblly=41.pt,bburx=416.pt,bbury=416.pt}}}
\caption[]{PUEO-KIR $H$-band image of APM~08279+5255. The correction is
performed on the object itself ($R$~=~15.2). The total magnitude is
$H$~=~12.6~$\pm$~0.1 and the double nature of the object is apparent: the
separation of the two components is 0$\farcs$35~$\pm$~0$\farcs$02 and the
intensity ratio $I_{\rm north}$/$I_{\rm south}$~=~1.21$\pm$0.25.}
\label{apm}
\end{figure}
We observed APM~08279+5255 using the CFHT AOB and
the near-IR camera KIR on May 7, 1998 at high airmass ($sec(z)=1.27$). The
seeing was 0$\farcs$8 and the AOB correction was determined from the source
itself. Four two-minutes images of the quasar were obtained in each quadrant
of the detector. A background image was formed by median-averaging these
frames having excised the object from each frame. Each of the four images were
flat-fielded by the normalized, dark-subtracted background frame. The images
were then aligned and added together yielding the final image of resolution
FWHM $\approx$ 0$\farcs$3 shown in Fig.~\ref{apm}. It is apparent from this and
each of the individual images that the object of total magnitude
$H$~=~12.6~$\pm$~0.1 is double. This observation strongly supports the lensing
interpretation for the source put forward by \citet{1998ApJ...505..529I}
from deconvolution of an $R$-band image. More
support for the lensing hypothesis comes from the uniformity of the quasar
spectrum as a function of spatial position in the image obtained with the
integral field spectrograph OASIS (see next Subsection).

We have used the point-spread function of a star of similar magnitude
subsequently observed to subtract the two components. This procedure leaves an
excess of magnitude $H$~$>$~16.5 between the two images. The origin and
reality of this excess is unclear: the stellar PSF may not be a perfect fit to
the quasar PSF due to variations in weather conditions and the nearly equal
flux of the two components could have led to a non-optimum AOB correction. We
have therefore performed a fit to the quasar double image minimizing the
chi-square between the data and a model PSF. The latter is composed of two
2D-gaussian functions assumed to reproduce the typical core plus halo
structure of the PSF in the case of small Strehl ratios \citep[see
e.g.][]{1996Msngr..83...14P}. The residual emission is of comparable strength
to that in the stellar-PSF subtraction, but has no preferred location. Note
that the residual is more than four magnitudes weaker than the quasar itself.
The procedure should therefore give a reliable, although approximate,
intensity ratio. From this fit, we derive that the separation between the two
images is 0$\farcs$35~$\pm$~0$\farcs$02 and the intensity ratio
$I_{\rm north}$/$I_{\rm south}$ = 1.21$\pm$0.25. The core and halo FWHM are
0$\farcs$26 and 0$\farcs$64 respectively and the intensity ratio
$I_{\rm core}$/$I_{\rm halo}$~$\approx$~3.7.

We have searched for additional images in a 18$\arcsec\times$18$\arcsec$
field. We have detected only one non-resolved object 10$\arcsec$ south-east of
the lens with a 5$\sigma$ magnitude of $H$~=~21.3.
\subsection{Integral field spectroscopy}
We used the new integral field spectrograph Optically Adaptative System for
Imaging Spectroscopy (OASIS) at the F/8 Cassegrain focus of the CFHT on March
29, 1998 to observe the 15$\arcsec\times$12$\arcsec$ field around the quasar.
The AOB correction could not be applied because of poor weather conditions;
the mean seeing was 1$\farcs$5 FWHM. Moreover, the sky was cloudy during the
exposure making the absolute photometry uncertain. Due to the position of the
object on the sky, a single 2700~s exposure was obtained. The O300 grism was
used in the wavelength range 5600-6200~\AA . The spectral resolution FWHM was
5.7~\AA\ as measured on the Neon arc lines used for wavelength calibration,
while the spatial scale was 0$\farcs$41 per lenslet.

The data reduction was performed using the dedicated software XOasis developed
by the Observatoire de Lyon (France). A set of $\sim$ 1100 separate spectra
were extracted from the CCD frame and rearranged in a 3D-datacube
($\alpha$,$\delta$,$\lambda$). During this process, an algorithm is used to
compute the lens array characteristics and the distortion coefficients of the
spectrograph optics. Wavelength and flat-field calibrations were consistently
done with the ``GUMBALL'' calibration system which allows the light from the
calibration source to follow the same optical path as that from the object.
Cosmic ray impacts were searched for in both the spatial and spectral
directions and removed.

\begin{figure}[t]
\flushleft{\vbox{
\psfig{figure=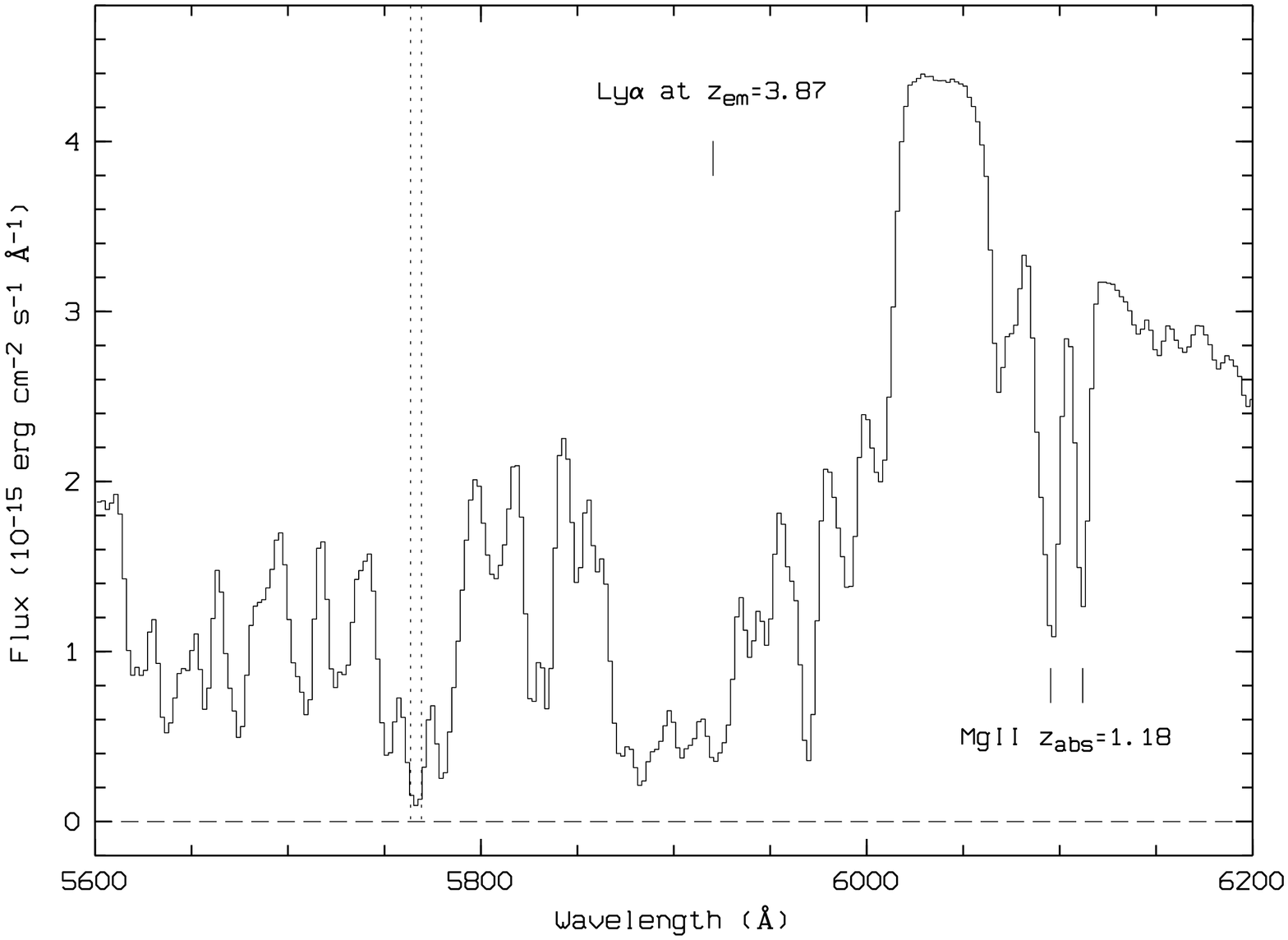,width=8.1cm,angle=-90.,clip=,bbllx=53.pt,bblly=78.pt,bburx=512.pt,bbury=788.pt}
\psfig{figure=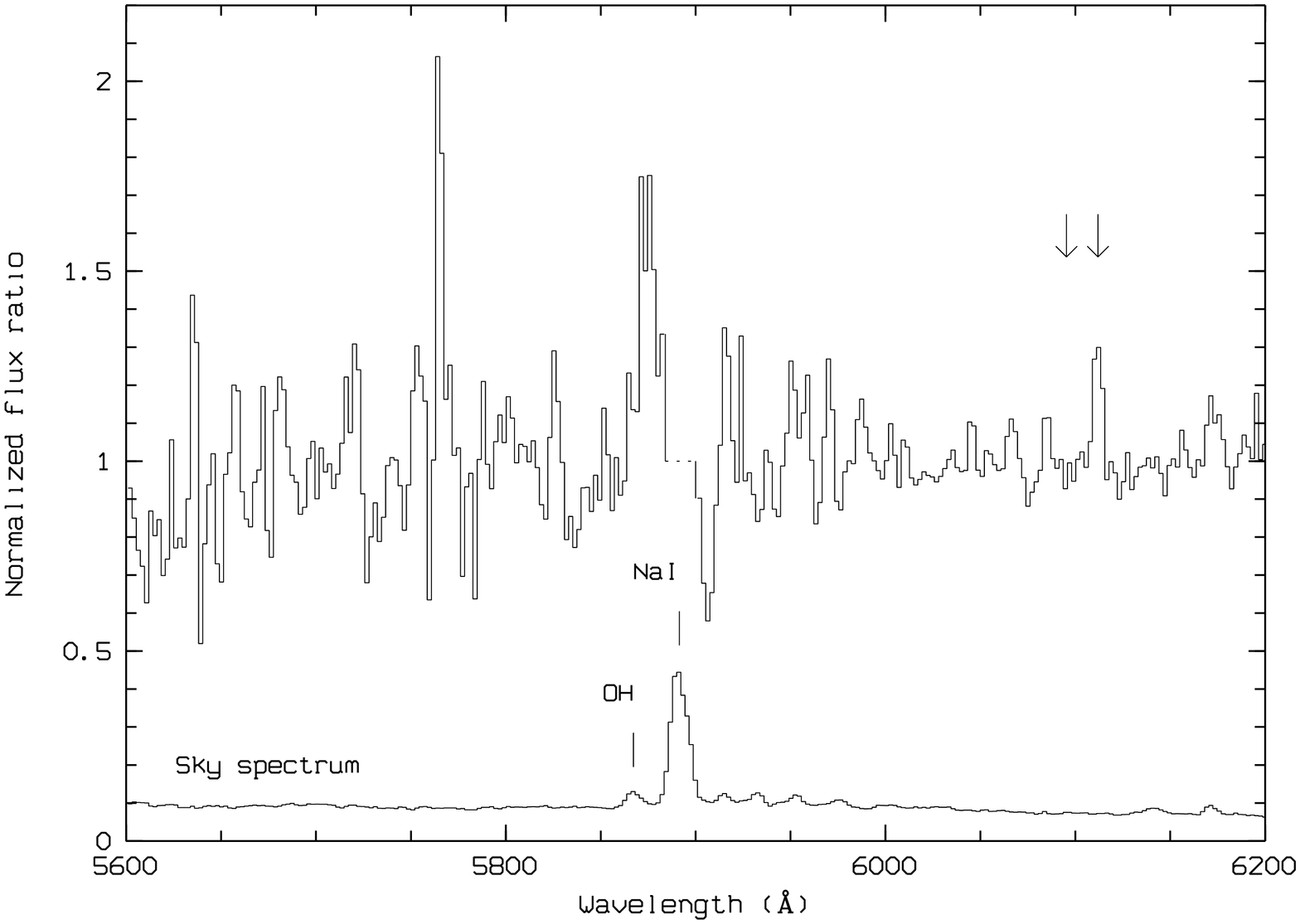,width=8.1cm,angle=-90.,clip=,bbllx=55.pt,bblly=78.pt,bburx=558.pt,bbury=788.pt}}}
\caption[]{Upper panel: Optical spectrum of the quasar APM~08279+5255 from
integral field spectroscopy. Due to cloudy conditions, the flux calibration is
uncertain by a factor 1.5. The dashed vertical lines indicate the position
and width of the filter used to reconstruct the narrow-band image shown in
Fig.~3. Lower panel: ratio of the north-eastern to the south-western spectra
(see text for details). There is no systematic difference between the two
spectra except in the Mg~{\sc ii}$\lambda$2803 absorption line at
$z_{\rm abs}$~=~1.18.}
\label{apmspe}
\end{figure}

The integrated spectrum of the quasar is shown in Fig.~\ref{apmspe}. The blue
component of the \ion{Mg}{ii} absorption doublet at 6095~\AA\ ($z_{\rm
abs}$~=~1.1796) is broader than its red counterpart. This is partly a
consequence of blending of this line with the \ion{N}{v}$\lambda$1242
counterpart of a \ion{N}{v}$\lambda$1238 line seen in absorption at $\approx$
6072~\AA .
However as the \ion{N}{v}$\lambda$1242 line cannot be stronger than the
\ion{N}{v}$\lambda$1238 line, the \ion{Mg}{ii} doublet must be multiple to
reproduce the absorption features. A fit to the doublet reveals the presence
of two components separated by $\sim$ 150~km~s$^{-1}$ and of column densities
log~$N$(\ion{Mg}{ii})~=~14.3 and 13.5 respectively. The redshift of the above
\ion{N}{v} doublet, $z_{\rm abs}$~=~3.9014, indicates that part of the gas is
falling towards the quasar with velocities larger than 1500~km~s$^{-1}$.

Since the relative positions of the two images of the quasar are known from
adaptive optics imaging, we have constructed two separate spectra (one for
each quasar image) by selecting spectra from the lenslets that are
predominantly illuminated by one of the images.
The seeing during the exposure was 1$\farcs$5 and the
distance between the two images is only 0$\farcs$35, this extraction must
therefore be performed with care to lead significative results.
We consider the superposition of two gaussian functions with
FWHM~=~1$\farcs$5, intensity ratio 1.2 and separation 0$\farcs$35. At a
distance of 1$\farcs$0 from the maximum, and in the direction joining their
center, the flux of one of the components always dominates the flux of the
other one by a factor larger than 2. This factor is even better in the
north-east since the north-eastern component is stronger. We have added the
spectra of 14 and 12 lenslets at a distance larger than 1$\farcs$2 from the
maximum (taking into account the lenslet size) and inside a radius of
0$\farcs$8 in the north-east and south-west directions respectively. To avoid
possible systematic errors due to the discrete 2D-spectroscopy, we have
performed the same extraction on the standard star that was observed just
after the science exposure. The ratio of the north-eastern to the
south-western spectra corrected from the standard star is shown in
Fig.~\ref{apmspe}. 
In the red, $\sigma$~=~0.1, and it is apparent
that there is no significant difference in the emission line range except at
6112~\AA\ (3$\sigma$ deviation). Interesting enough, this is the wavelength 
of the
\ion{Mg}{ii}$\lambda$2803 absorption line at $z_{\rm abs}$~=~1.18 suggesting
that the doublet ratio is weaker (and so is the column density) in
front of the northern image, indicative of small-scale
spatial variations of the column densities in this strong system. In the
Ly$\alpha$ forest, $\sigma$~=~0.25. 
However, it is apparent that there is no difference in
the strong broad absorptions just blueward the Ly$\alpha$ emission
line. This altogether is a good indication that
the quasar is lensed. Indeed, if the two quasars were different, we would
expect to see strong differences in the BAL as it is the case for
HS~1216+5032A,B
\citep{1996A&A...308L..25H}. BALs are expected to arise in gas ejected by the
quasar with rapid spatial variations \citep{1998ApJ...496..177M}.
There is an apparently strong deviation at $\lambda$5885 but this
is due to the presence of the very strong sky emission line. There may
be a significant difference at $\lambda$5874 and $\lambda$5766 but it must be 
noticed that the uncertainty at these wavelengths is very large 
since these two wavelengths correspond to minima in the spectrum.

\begin{figure*}[t]
\flushleft{\hbox{
\psfig{figure=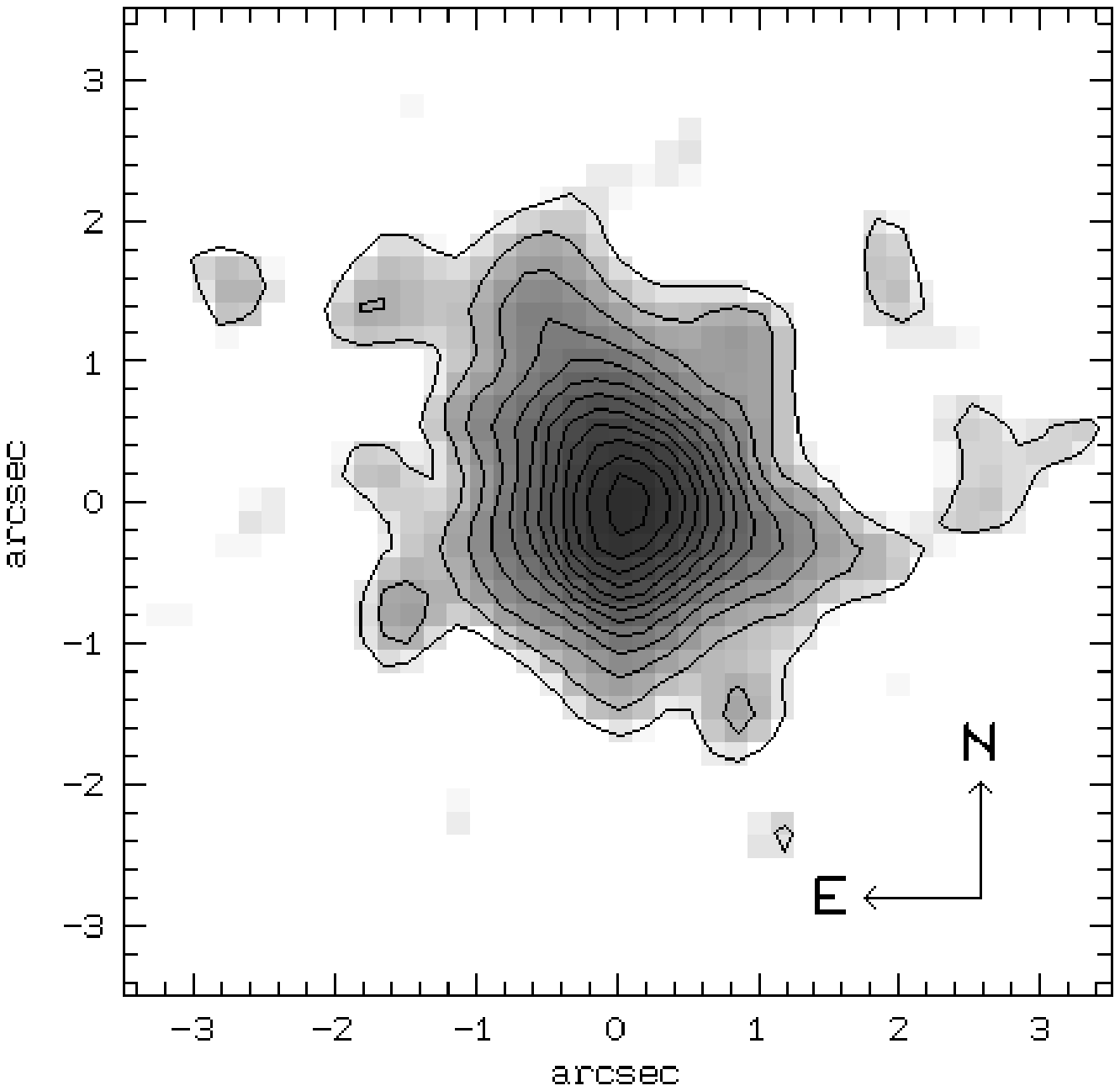,width=8.1cm,clip=,bbllx=89.pt,bblly=331.pt,bburx=473.pt,bbury=706.pt}
\psfig{figure=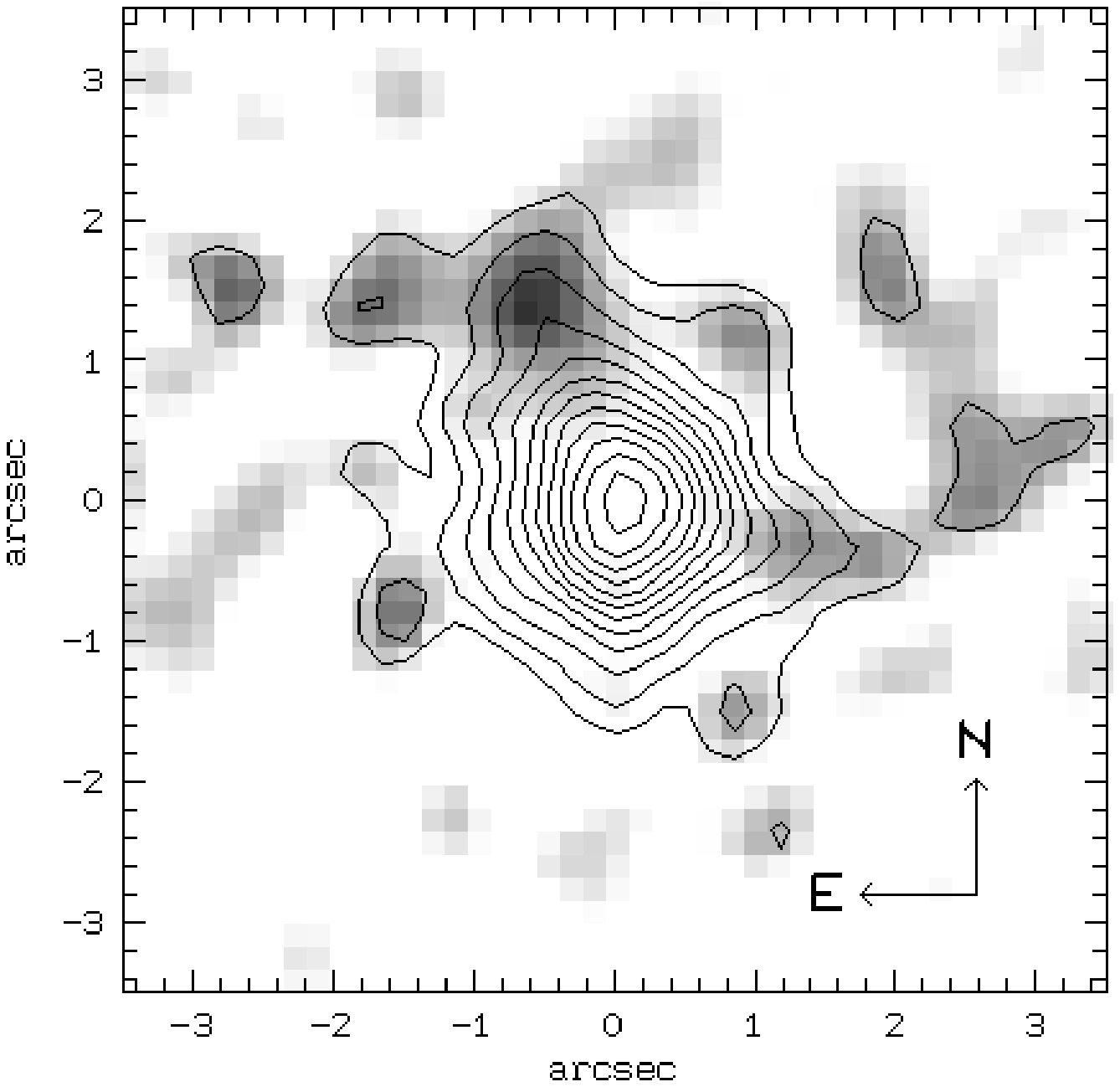,width=8.1cm,clip=,bbllx=89.pt,bblly=331.pt,bburx=473.pt,bbury=706.pt}}}
\caption[]{Narrow-band image of the field surrounding APM~08279+5255. The
central wavelength and the width of the filter are 5766.4~\AA\ and 5.7~\AA\
respectively, centered on the deepest Ly$\alpha$ absorption seen in the forest
of the quasar spectrum. The left panel corresponds to the image reconstructed
from integral field spectroscopy. In the right panel, the quasar image has
been subtracted.
The magnitude of the 3$\sigma$ excess (the cut levels have been adjusted to
enhance the contrast) seen 1$\farcs$5 to the north-east of the quasar is
$R$~$\sim$~20.6 if its spectrum is assumed to be flat over the $R$-band.}
\label{pixblack}
\end{figure*}
%
Using the flux-calibrated object datacube, narrow-band images of the field can
be reconstructed at any wavelength within the observed range to search for
line-emitting objects. Without any quasar subtraction, we find no such object
to a limit of 6$\times$10$^{-17}$ erg cm$^{-2}$ s$^{-1}$. The
redshift range probed by our observations is 0.503-0.664 for
[\ion{O}{ii}]$\lambda$3727 and 3.609-4.103 for \ion{H}{i} Ly$\alpha$.
Unfortunately, the current observations do not cover the redshifts of either
the candidate damped system at $z_{\rm abs}$~= 3.07
\citep{1998ApJ...505..529I} or the strong \ion{Mg}{ii} system at $z_{\rm
abs}$~=~1.18. The quasar, although attenuated by a large factor in the
Ly$\alpha$ forest, is still bright in most of the images however. In
order to detect objects lying in projection close to the quasar, we 
have performed a subtraction of the residual light
using the image of the quasar integrated over the whole
wavelength interval and then scaled to the residual intensity in the
narrow-band filter. This procedure takes into account the exact shape
of the double image and, because of the scaling operation,
subtracts very few of an hypothetical flat-spectrum or
line-emitting object except at wavelengths where the quasar is at its
maximum. This prevents however to detect sources with the same spectrum 
as the quasar.

No prominent line-emitting object is detected but, when
performing the QSO subtraction, residual intensity is seen in a
region 1$\farcs$5 away from the quasar to the north-east.
To check whether this residual is an artifact of the QSO subtraction or not,
we can take advantage of the fact that the Ly$\alpha$ line at 5766.4~\AA\ is
nearly black to construct an image where the quasar is at its
minimum\footnote{Note that the non-zero level in this spectral bin (see
Fig.~\ref{apmspe}) is due to the low resolution of the spectrum. Profile
fitting of the line shows that even for a column density as large as
log~$N$(\ion{H}{i})~$\sim$~19.4 and a Doppler parameter as small as
b~$\sim$~10~km~s$^{-1}$, the spectrum is not expected to go to the zero
level.}. The image in a narrow spectral band centered at 5766.4~\AA\ and of
width 5.7~\AA , corresponding to the spectral resolution element (FWHM), is
shown in Fig.~\ref{pixblack} (left panel). 
The image obtained after subtraction of the QSO is shown in the same figure
(right panel). If the excess seen 1$\farcs$5
north-east of the quasar has a flat-continuum spectrum, then the $R$ magnitude
of the object is of the order of 20.6. To increase the signal-to-noise ratio
in the image, we have applied the same technique for all the Ly$\alpha$
absorption lines. An excess is seen at the same location but with magnitude
$R$~$\sim$~21 if the spectrum is assumed to be from a flat-continuum source.
Since the excess is only measured at the 3$\sigma$ level and is not seen in
the $H$-band adaptive optics image, this cannot be 
considered as a firm detection and additional data is needed to confirm the 
presence of an object. 
%
\section{Conclusion}
From the fit of the $H$-band adaptive optics image of APM~08279 +5255, we have
confirmed that this quasar is double with separation between the two sources
0$\farcs$35~$\pm$~0$\farcs$02 and an intensity ratio $I_{\rm north}$/$I_{\rm
south}$~=~1.21$\pm$0.25. From integral field spectroscopy, we derive that
there is no systematic difference between the spectra of the 
north-eastern and south-western sources except possibly in the
Mg~{\sc ii}$\lambda$2803 absorption line at $z_{\rm abs}$~=~1.18.
This is clear evidence that the quasar is gravitationally lensed.
For the given characteristics, a simple singular isothermal sphere model
implies an amplification factor of the order of 20 and a velocity dispersion
of 120~km~s$^{-1}$. This amplification factor implies an intrinsic luminosity
of the order of a few 10$^{14}$~L$_{\odot}$
\citep[see][]{1998ApJ...505..529I}. The velocity dispersion is consistent with
the velocity spread of an intervening \ion{Mg}{ii} absorption line system at
$z_{\rm abs}$~=~1.18.

An image separation of only 0$\farcs$35 implies that, either the lensing
object is very compact and in between the two images, or that the source is
close to a caustic of the deflection mapping. In the latter case, another
faint image should be seen elsewhere in the field \citep[see
e.g.][]{1980ApJ...241..507Y,1992grle.book.....S}. No other image is seen
within 10$\arcsec$ from the quasar down to $H$~=~21.3.
Narrow-band images centered on the strongest Ly$\alpha$ absorption lines in
the forest and reconstructed from integral field spectroscopy show a 3$\sigma$
excess of light of magnitude $R$~$\sim$~21 (assuming a flat-continuum
spectrum) at 1$\farcs$5 north-east of the quasar. However the reality of the
source is not secure given the poor weather conditions during the
observations.

It is intriguing that the broad absorption spread over more than 50~\AA\
around 5900~\AA\ does not go to the zero level. This indicates that 
it is not continuous and
should split into numerous narrower components at high resolution.
The  presence of \ion{N}{v} absorption at 6072~\AA\ indicates the presence of
associated systems with redshifts larger than the emission redshift of the
quasar by at least 1500 km~s$^{-1}$.
\acknowledgements{We are grateful to Emmanuel P\'econtal, Eric Emsellem,
Pierre Martin and the OASIS team for their efficient support, and to Pierre
Couturier for his help.}
\bibliographystyle{cedric}
\bibliography{apm}
\end{document}